\documentclass[aps,prd,preprint,groupedaddress,longbibliography]{revtex4-1}

\usepackage{amsmath,amssymb,amsfonts}
\usepackage[T1]{fontenc}
\usepackage{graphicx}
\usepackage{adjustbox}
\usepackage{array}
\newcolumntype{L}{>{\centering\arraybackslash}p{3.2cm}}
\usepackage{hyperref}
\usepackage{multirow}

\begin{document}

\title{Invisible decays of neutral hadrons}


\author{Wanpeng Tan}
\email[]{wtan@nd.edu}
\affiliation{Department of Physics, Institute for Structure and Nuclear Astrophysics (ISNAP), and Joint Institute for Nuclear Astrophysics - Center for the Evolution of Elements (JINA-CEE), University of Notre Dame, Notre Dame, Indiana 46556, USA}

\date{\today}

\begin{abstract}
Invisible decays of neutral hadrons are evaluated as ordinary-mirror particle oscillations using the newly developed mirror matter model. Assuming equivalence of the $CP$ violation and mirror symmetry breaking scales for neutral kaon oscillations, rather precise values of the mirror matter model parameters are predicted for such ordinary-mirror particle oscillations. Not only do these parameter values satisfy the cosmological constraints, but they can also be used to precisely determine the oscillation or invisible decay rates of neutral hadrons. In particular, invisible decay branching fractions for relatively long-lived hadrons such as $K^0_L$, $K^0_S$, $\Lambda^0$, and $\Xi^0$ due to such oscillations are calculated to be $9.9\times 10^{-6}$, $1.8\times 10^{-6}$, $4.4\times 10^{-7}$, and $3.6\times 10^{-8}$, respectively. These significant invisible decays are readily detectable at existing accelerator facilities.
\end{abstract}


\maketitle

\section{Introduction\label{intro}}

Searching for new physics beyond the Standard Model (BSM), such as the puzzle of dark matter, has stimulated studies on invisible decays of heavy hadrons like $D^0$ \cite{bellecollaboration2017}, $B^0$ \cite{bellecollaboration2012}, and Higgs particles \cite{atlascollaboration2019,sirunyan2019}. But the attention to possible invisible decays in hadrons made of light quarks \cite{besiiicollaboration2018} has been much more scarce than one would think. Early measurements of invisible decays on short-lived light mesons of $\pi^0$ \cite{e949collaboration2005} and $\eta$ ($\eta'$) \cite{besiiicollaboration2013} were attempted. Search for invisible decays of $\omega$ and $\phi$ were also conducted by the BESIII collaboration \cite{besiiicollaboration2018}. However, no experiment has been conducted for invisible decays of relatively long-lived neutral hadrons like $K^0$ \cite{gninenko2015}, which present the most significant $CP$-violation effect to our knowledge. Insufficient experimental work in light hadrons may miss the discovery of BSM physics that could very well be revealed in their invisible decays.

Invisible decays via neutrino emission $\bar{\nu}\nu$ were calculated for heavy mesons of $B^0$ and $D^0$ with negligible branching fractions (on the order of $10^{-24}$ or lower) \cite{badin2010} while the four-neutrino ($\bar{\nu}\nu\bar{\nu}\nu$) decay channel can be much more enhanced due to lack of helicity suppression with a still low branching fraction (on the order of $10^{-15}$ for $B^0$) \cite{bhattacharya2019}. Calculations for light meson decays via $\bar{\nu}\nu$ and $\bar{\nu}\nu\bar{\nu}\nu$ have also shown very low branching ratios \cite{gao2018}. Therefore, these invisible neutrino decay channels will not hide new physics if a much larger branching fraction (e.g., $\gg 10^{-13}$) due to BSM physics is predicted. Indeed, unexpectedly large rates of ordinary-mirror particle oscillations of long-lived neutral hadrons predicted under the newly developed mirror-matter model \cite{tan2019,tan2019c,tan2019d,tan2019e} can manifest as significant invisible decays of these particles.

In this work, we will present a rather exact model of neutral hadron-mirror hadron oscillations inferred from the similarity between $CP$ violation and mirror symmetry breaking. Assuming equivalence of the $CP$-violation and mirror symmetry breaking scales, one can pin down the model parameters very precisely leading to precise predictions of large invisible decay branching fractions for long-lived neutral hadrons that can be detected at existing and planned facilities.

\section{Mirror Matter Model\label{sm3}}

The mirror matter idea was originated from the discovery of parity violation in the weak interaction by Lee and Yang \cite{lee1956}. It has subsequently been developed into an intriguing mirror-matter theory by various efforts \cite{kobzarev1966,blinnikov1983,kolb1985,hodges1993,berezhiani2004,berezhiani2006,cui2012,foot2014}. The general picture of the theory is that a parallel sector of mirror particles (denoted by the prime symbol below) exists as an almost exact mirrored copy of the known ordinary particles and the two worlds can only interact with each other gravitationally. Nevertheless, many of previous models \cite{berezhiani2004,berezhiani2006,cui2012,foot2014} attempted to add some explicit feeble interaction between the two sectors. On the contrary, in the newly proposed mirror-matter model \cite{tan2019,tan2019c,tan2019d,tan2019e}, no explicit cross-sector interaction is introduced, namely, the two parallel sectors share nothing but the same gravity before the mirror symmetry is spontaneously broken.

The new mirror-matter model has been applied to consistently and quantitatively solve a wide variety of puzzles in physics and cosmology: dark matter and neutron lifetime anomaly \cite{tan2019}, evolution of stars \cite{tan2019a}, matter-antimatter imbalance \cite{tan2019c}, ultrahigh energy cosmic rays \cite{tan2019b}, unitarity of CKM \cite{tan2019d}, neutrinos and dark energy \cite{tan2019e}. The model has also been extended into a set of supersymmetric mirror models under dimensional evolution of spacetime to explain the arrow of time and big bang dynamics \cite{tan2020,tan2020a} and to understand the nature of black holes \cite{tan2020b}. Last but not least, various feasible laboratory experiments are proposed to test the new theory \cite{tan2019d}.

An immediate result of this model is the probability of ordinary-mirror neutral particle oscillations in vacuum \cite{tan2019} assuming no decay or a much shorter time scale than its decay lifetime,
\begin{equation}\label{eq_prob}
P(t) = \sin^2(2\theta) \sin^2(\frac{1}{2}\Delta t)
\end{equation}
where $\theta$ is the mixing angle, $\sin^2(2\theta)$ denotes the mixing strength (on the order of $10^{-6} \text{--} 10^{-4}$ for the most significant $n-n'$ and $K^0-K^{0'}$ oscillations \cite{tan2019}), $t$ is the proper propagation time, and $\Delta$ is the small mass difference of the two mass eigenstates (on the order of $10^{-6}$ eV for both $K^0-K^{0'}$ and $n-n'$ \cite{tan2019}). Here natural units are used for simplicity.

Under the new model, the spontaneous mirror symmetry breaking results in one extended quark mixing matrix as follows \cite{tan2019d,tan2019e},
\begin{equation} \label{eq_qmix}
V_\text{qmix} = 
\begin{pmatrix}
\multicolumn{3}{c}{\multirow{3}{*}{\Large $V_\text{CKM}$}} & \vdots & V_{uu'} & & \\
\multicolumn{3}{c}{} & \vdots & & V_{cc'} & \\
\multicolumn{3}{c}{} & \vdots & & & V_{tt'} \\
\cdots & \cdots & \cdots & \cdots & \cdots & \cdots & \cdots \\
V_{dd'} & & & \vdots & \multicolumn{3}{c}{\multirow{3}{*}{\Large $V'_\text{CKM}$}} \\
& V_{ss'} & & \vdots & \multicolumn{3}{c}{} \\
& & V_{bb'} & \vdots & \multicolumn{3}{c}{}
\end{pmatrix}
\end{equation}
where the ordinary $3\times3$ CKM matrix $V_\text{CKM}$ and its mirror counterpart $V'_\text{CKM}$ are no longer unitary. 
Using the results from neutron lifetime measurements via the ``beam'' approach or superallowed nuclear $0^+ \rightarrow 0^+$ decays for the matrix element $V_{ud}$, the deviation from unitarity of the ordinary CKM matrix was demonstrated at a significance level of $>5\sigma$ \cite{tan2019d}.
The missing strength from unitarity is supplied by quark-mirror quark mixing elements $V_{qq'}$. The unitarity condition for the first row of the extended matrix can then be written as,
\begin{equation}\label{eq_unitary}
|V_{ud}|^2+ |V_{us}|^2+ |V_{ub}|^2+ |V_{uu'}|^2=1.
\end{equation}

The matrix element $|V_{ub}| = 0.00394(36)$ \cite{particledatagroup2018} contributes little to the unitarity. The best direct constraint on matrix element $V_{us}$ can be set from measurements of meson decays as $|V_{us}| = 0.22333(60)$ under the new scenario \cite{moulson2017,tan2019d}. Using the neutron $\beta$-decay lifetime of $\tau_n = 888.0\pm 2.0$ s from the averaged ``beam'' values \cite{byrne1996,yue2013}, we can obtain the matrix elemment $|V_{ud}| = 0.9684(12)$ \cite{tan2019d}.
As shown in Table \ref{tab_1}, the unitarity requirement of Eq. (\ref{eq_unitary}) then results in $|V_{uu'}| \simeq 0.11(1)$ with the known values of the above matrix elements.

\begin{table}
\caption{\label{tab_1} Adopted or predicted element values of the extended unitary mixing matrix in Eq. (\ref{eq_qmix}) under the new mirror-matter model are listed. In particular, $V_{ud}$ is determined from the ``beam'' lifetime approach \cite{byrne1996,yue2013} while $V_{us}$ is taken from the semileptonic $K_{l3}$ decay measurements \cite{moulson2017}  and $V_{ub}$ is from Ref. \cite{particledatagroup2018}. The cross-sector mirror mixing elements are predicted using the unitarity and the $n-n'$ and $K^0-K^{0'}$ mixing strengths as discussed in the text.}
\begin{ruledtabular}
\begin{tabular}{c c c c c c c c c}
$|V_{ud}|$ & $|V_{us}|$ & $|V_{ub}|$ & $|V_{uu'}|$ & $|V_{dd'}|$ & $|V_{ss'}|$ & $|V_{cc'}|$ & $|V_{bb'}|$ & $|V_{tt'}|$ \\
\hline
0.9684(12) & 0.22333(60) & 0.00394(36) & 0.11(1) & 0.063 & 0.018 & 0.005 & 0.0014 & 0.0004 \\
\end{tabular}
\end{ruledtabular}
\end{table}

To estimate the strengths of the other quark-mirror quark mixing elements, we can apply the following relationship between the mixing strength of a neutral hadron and the mirror-mixing matrix elements for its corresponding constituent quarks,
\begin{equation}\label{eq_mixv}
\sin^2(2\theta) \simeq \prod_{i} |2V_{q_iq'_i}|^2
\end{equation}
where the mixing angle $\theta$ is assumed to be small. Using Eq. (\ref{eq_mixv}), more mirror-mixing elements can be estimated from the known mixing strengths of ordinary-mirror hadron oscillations. For example, the $n-n'$ mixing strength $\sin^2(2\theta_{nn'}) = |2V_{uu'}|^2|2V_{dd'}|^4 \sim 2\times 10^{-5}$ leads to $|V_{dd'}| \sim 0.071$ \cite{tan2019d}. The study of $K^0-K^{0'}$ oscillations in the early universe for the origin of baryon asymmetry \cite{tan2019c} supports the mixing strength $\sin^2(2\theta_{KK'}) = |2V_{dd'}|^2|2V_{ss'}|^2 \sim 10^{-4}$ resulting in $|V_{ss'}| \sim 0.035$ \cite{tan2019d}. Unfortunately, these mixing strengths without better experimental constraints still carry large uncertainties and we shall present a better way to pin down these parameters precisely.

\section{$CP$ violation and mirror symmetry breaking\label{cp}}

By analyzing the published neutron lifetime measurements via the ``bottle'' approach, a probable range of $8\times10^{-6}$ -- $4\times10^{-5}$ for the $n-n'$ mixing strength is obtained \cite{tan2019}. According to a recent simulation result for the UCN$\tau$ setup \cite{pattie2018}, the mean free flight time for neutrons in the trap is estimated to be $0.33\pm 0.08$ s \cite{liu2019note}. This gives an estimate of $\sin^2(2\theta) = 8.7\pm 2.8\times 10^{-6}$ for $n-n'$ oscillations, which is effectively a lower limit on the mixing strength as higher energy neutrons tend to disappear more easily in the trap via $n-n'$ oscillations and the effect was not taken into account in the simulation. On the other hand, an anomaly of $707\pm20$ s in neutron lifetime measured in a $^4$He-filled magnetic trap at NIST \cite{huffer2017} in combination with a new simulation study \cite{coakley2020note} presents another estimate of $\sin^2(2\theta) \simeq 2\times 10^{-5}$. This in fact provides an upper limit of the mixing strength since there could exist other unaccounted processes (e.g., from $^3$He contamination) that may contribute to the apparently short lifetime value. For such a mixing strength range of $0.9 \text{--} 2\times 10^{-5}$, one can obtain a range of $|V_{dd'}|=0.058 \text{--} 0.071$ using Eq. (\ref{eq_mixv}) as discussed above. Without further magnetic trap measurements, however, could we constrain $V_{dd'}$ better? One way is to infer it from the mass splitting parameter as presented below.

For ordinary $CP$-violating neutral kaon oscillations, its mixing and mass splitting parameters have been well measured.
It is natural to postulate that the $CP$ violation scale is the same as the mirror oscillation scale, in particular, for the case of neutral kaon oscillations \cite{tan2019e}. This may very well be the case if we consider $CP$ violation as a direct result of spontaneous mirror symmetry breaking at staged quark condensation \cite{tan2019e}. Therefore, we assume that the mixing angle and mass splitting scale are the same for both symmetry breakings as follows,
\begin{eqnarray}
\Delta_{K^{0}K^{0'}} = \Delta_{K^{0}_LK^{0}_S} = 3.484(6)\times 10^{-6}\ \text{eV} \nonumber \\
\sin\theta_{K^{0}K^{0'}} = \sin\theta_{K^{0}_LK^{0}_S} = 2.228(11)\times10^{-3} \label{eq_cpsame}
\end{eqnarray}
where the well-measured $CP$ violation parameter values are taken from the PDG compilation \cite{particledatagroup2018}.

As a particle's mass originates from the same Higgs mechanism, we assume that the mass splitting parameter scales as the particle's mass, i.e., the relative mass splitting scale of $\Delta/m \simeq 7\times 10^{-15}$ derived from Eq. (\ref{eq_cpsame}) is constant under the mechanism of spontaneous mirror symmetry breaking \cite{tan2019e}. Therefore, we can obtain a rather precise value of $\Delta_{nn'}=6.578\times 10^{-6}$ eV for $n-n'$ oscillations from the well known kaon data, which is similar to the value estimated under the cosmological consideration in the new mirror matter model \cite{tan2019,tan2019c}.

For consistent origin of dark (mirror) matter and baryon asymmetry of the universe, we can obtain an approximate relationship of the mixing strength and mass splitting parameters using the dark-to-baryon density ratio of 5.4 \cite{tan2019c},
\begin{equation}\label{eq_dark5.4}
\sin^2(2\theta_{nn'}) = \left( \frac{4.4\times 10^{-14}\ \text{eV}}{\Delta_{nn'}}\right)^{0.6}
\end{equation}
which gives $\sin^2(2\theta_{nn'})=1.25\times 10^{-5}$ for $\Delta_{nn'}=6.578\times 10^{-6}$ eV assuming equivalence of the $CP$ violation and mirror symmetry breaking scales. Amazingly, this mixing strength value falls right within the range of $0.9 \text{--} 2\times 10^{-5}$ constrained by the above-discussed neutron lifetime measurements. Using Eq. (\ref{eq_mixv}), the $n-n'$ mixing strength $\sin^2(2\theta_{nn'}) = |2V_{uu'}|^2|2V_{dd'}|^4 \simeq 1.25\times 10^{-5}$ then leads to $|V_{dd'}| \simeq 0.063$ as shown in Table \ref{tab_1}.

Under the equivalence of the $CP$ and mirror symmetry breaking scales in Eq. (\ref{eq_cpsame}), we can also find the following relationship between the mixing strength and mass splitting parameters for $K^0-K^{0'}$ oscillations,
\begin{equation}\label{eq_kkrel}
\sin^2(2\theta_{K^0K^{0'}})\Delta_{K^0K^{0'}}^2 = 2.41\times 10^{-16}\ \text{eV}^2
\end{equation}
which is remarkably close to what is required to account for the baryon asymmetry of the universe using $K^0-K^{0'}$ oscillations \cite{tan2019c}. From its mixing strength of $\sin^2(2\theta_{K^0K^{0'}})\simeq 2\times 10^{-5}$ and Eq.(\ref{eq_mixv}), one can derive another mirror mixing matrix element of $|V_{ss'}| \simeq 0.018$ as shown in Table \ref{tab_1}.

There is no adequate data to directly determine the mirror mixing matrix elements for heavier quarks. For simplicity, we can assume that these matrix elements follow a geometric sequence with common ratio of $r=|V_{ss'}/V_{dd'}|\simeq 0.28$ corresponding to the quark hierarchy, e.g., $|V_{cc'}| = r |V_{ss'}|$. Then we can easily calculate these matrix elements for heavier c-, b-, and t-quarks as shown in Table \ref{tab_1}. This will provide a convenient way to estimate the invisible decay branching ratios of heavier neutral hadrons as discussed below. Obviously, the uncertainty goes much larger when heavier quarks are involved.

\section{Invisible decays of neutral hadrons\label{inv}}

For a neutral hadron $h$ during the process of free decay, we can calculate its $h-h'$ oscillation rate using Eq. (\ref{eq_prob}) as follows,
\begin{eqnarray}\label{eq_osc}
\lambda_{hh'} &=& \frac{1}{\tau^2}\int_0^{\infty} P(t) \exp(-t/\tau) dt \nonumber \\
&=& \frac{1}{2\tau} \sin^2(2\theta) \frac{(\Delta \tau)^2}{1+(\Delta \tau)^2}
\end{eqnarray}
where $\tau$ is the hadron's mean lifetime and $\Delta$ is the mass splitting parameter of $h-h'$ oscillations. The corresponding invisible decay branching ratio due to $h-h'$ oscillations can then be written as,
\begin{equation} \label{eq_inv}
B_\text{inv} = \frac{1}{2} \sin^2(2\theta) \frac{(\Delta \tau)^2}{1+(\Delta \tau)^2}.
\end{equation}

It is easy to see in Eq. (\ref{eq_inv}) that a neutral hadron has to be relatively long-lived, i.e., with lifetime of $\tau \gtrsim 1/\Delta$, for its invisible decay branching ratio to be significant. With the universal relative mass splitting scale of $\Delta/m \simeq 7\times 10^{-15}$ as discussed above, one can calculate mass splitting parameters of long-lived hadrons as listed in Table \ref{tab_2}. The longest-lived neutral hadrons are neutrons and we can obtain $B_\text{inv}(n)\simeq 6.3\times 10^{-6}$ using the above-discussed parameters. Such an effect is greatly amplified by wall scattering and readily observed for neutrons confined in a trap \cite{tan2019,tan2019d}.

\begin{table}
\caption{\label{tab_2} Branching fractions of invisible decays via ordinary-mirror hadron oscillations and corresponding mirror symmetry breaking parameters for long-lived neutral hadrons are listed. Lifetime values are taken from the PDG compilation \cite{particledatagroup2018}.}
\begin{ruledtabular}
\begin{tabular}{c c c c c c c}
Hadrons & $K^0_L$ & $K^0_S$ & $\Lambda^0$ & $\Xi^0$ & $D^0$ & $B^0$ \\
\hline
$\Delta_{hh'}$ [eV]& $3.484\times 10^{-6}$ & $3.484\times 10^{-6}$ & $7.8\times 10^{-6}$ & $9.2\times 10^{-6}$ & $1.3\times 10^{-5}$ & $3.7\times 10^{-5}$ \\
$\sin^2(2\theta_{hh'})$ & $2\times 10^{-5}$ & $2\times 10^{-5}$ & $9.8\times 10^{-7}$ & $7.6\times 10^{-8}$ & $4.8\times 10^{-6}$ & $1.2\times 10^{-7}$ \\
lifetime [s] & $5.116\times 10^{-8}$ & $8.954\times 10^{-11}$ & $2.632\times 10^{-10}$ & $2.9\times 10^{-10}$ & $4.101\times 10^{-13}$ & $1.519\times 10^{-12}$ \\
$B_{\text{inv}}$ & $9.9\times 10^{-6}$ & $1.8\times 10^{-6}$ & $4.4\times 10^{-7}$ & $3.6\times 10^{-8}$ & $1.6\times 10^{-10}$ & $4.4\times 10^{-10}$
\end{tabular}
\end{ruledtabular}
\end{table}

Using the rather precise values of the mirror mixing matrix elements listed in Table \ref{tab_1} under the assumption of equivalence of the $CP$ / mirror breaking scales, we can calculate the mirror mixing strengths and the corresponding invisible decay branching fractions of neutral hadrons with significant ordinary-mirror particle oscillations. As shown in Table \ref{tab_2}, the results for relatively long-lived neutral mesons and baryons are $B_\text{inv}(K^0_L) \simeq 9.9\times 10^{-6}$, $B_\text{inv}(K^0_S) \simeq 1.8\times 10^{-6}$, $B_\text{inv}(\Lambda^0) \simeq 4.4\times 10^{-7}$, $B_\text{inv}(\Xi^0) \simeq 3.6\times 10^{-8}$, $B_\text{inv}(D^0) \simeq 1.6\times 10^{-10}$, and $B_\text{inv}(B^0) \simeq 4.4\times 10^{-10}$. In particular, the large branching fractions for $K^0_L$, $K^0_S$, $\Lambda^0$, and $\Xi^0$ should be detectable at current and planned facilities. Note that the predicted invisible decay branching fractions for $K^0_L$ and $K^0_S$ are consistent with the indirect experimental upper limits of $<6.3\times 10^{-4}$ and $<1.1\times 10^{-4}$, respectively, set from existing visible decay data \cite{gninenko2015}.

For comparison, very short-lived neutral hadrons have negligible branching ratios due to the invisible oscillations. For example, $\pi^0$ with a lifetime of $8.52\times 10^{-17}$ s, though its mixing strength of $7.8\times 10^{-4}$ is quite large, has a very small $B_\text{inv} \simeq 6\times 10^{-18}$. The Higgs particle as a top-quark condensate \cite{tan2019e} can be estimated to have an even smaller $B_\text{inv} \sim 10^{-33}$.

Based on the known observational and experimental evidence, these large yet realistic estimates of the oscillation effects under the new mirror matter model may help motivate more experimental efforts in search of invisible decays of long-lived light neutral hadrons. New physics could be revealed in such experiments and it is also one of the promising laboratory approaches to test the unique predictions of the new mirror matter model \cite{tan2019d}.

\section{Further discussions\label{conclusion}}

Recent observation of anisotropy of the universe \cite{colin2019,migkas2020,wilczynska2020} could be well understood under the new framework of mirror matter theory, in particular, with the new supersymmetric mirror models \cite{tan2019e,tan2020}. Using the new result of a large dipole component of cosmic acceleration in a reanalysis of type Ia supernova data by Colin \textit{et al.} \cite{colin2019}, i.e., the monopole $q_m=-0.157$ and the dipole $q_d=-8.03$ of the cosmic deceleration parameter, we can further discuss its effects on the parameters of the new mirror matter model.

In this scenario, the monopole component $q_m$ dominates at large cosmic scales that should be defined by the universe contents of ordinary matter, dark (mirror) matter, and dark (vacuum) energy. Under the standard cosmic model $\Lambda$CDM assuming a flat universe, we can estimate that the content of dark energy would be $\Omega_\Lambda = (1-2q_m)/3 \simeq 44$\%. Assuming 5\% for the content of ordinary matter, dark (mirror) matter takes up about 51\%. Therefore, taking into account the possible cosmic anisotropy of Colin \textit{et al.} \cite{colin2019}, the dark-to-baryon density ratio is about 10:1, which seems to agree better with observation of galactic dark matter. This almost doubled ratio will modify the relationship between the $n-n'$ mass difference and its mixing strength slightly from Eq. (\ref{eq_dark5.4}) as follows,
\begin{equation}\label{eq_dark10}
\sin^2(2\theta_{nn'}) = \left( \frac{6.7\times 10^{-14}\ \text{eV}}{\Delta_{nn'}}\right)^{0.6}
\end{equation}
which leads to $\sin^2(2\theta_{nn'})=1.6\times 10^{-5}$ for $\Delta_{nn'}=6.578\times 10^{-6}$ eV. Intriguingly, this $n-n'$ mixing strength value is still within the experimental limits of $0.9 \text{--} 2\times 10^{-5}$. It modifies $n-n'$ oscillations slightly but has no effect on the invisible decay branching fractions of $K^0_L$, $K^0_S$, and $\Lambda^0$. The predictions on the other neutral hadrons are changed slightly to $B_\text{inv}(\Xi^0) \simeq 2.8\times 10^{-8}$, $B_\text{inv}(D^0) \simeq 1.1\times 10^{-10}$, and $B_\text{inv}(B^0) \simeq 2.7\times 10^{-10}$.

One caveat in the assumption of a universal relative mass splitting scale for mirror symmetry breaking could be that the role of the strong interaction may complicate the mass generation mechanism in hadrons. There may be some quark-dependent effects making the relative mass splitting scale not exactly constant. In the case of $D^0$, its mass splitting scale is $8.0^{+3.5}_{-3.7}\times 10^{-6}$ eV from a global fit of all experimental data by the HFLAV collaboration assuming $CP$ violation is negligible \cite{heavyflavoraveraginggroup2019}, which deviates just by a little over one sigma from the scale inferred from kaon oscillations as shown in Table \ref{tab_2}. This indicates that a constant relative mass splitting scale is at least a good approximation. For the $B^0$ system, on the other hand, the deviation is about one order of magnitude. But it could be a result of contamination from $B^0_s$.

\begin{acknowledgments}
I'd like to thank Hong-Jian He, Hai-Bo Li, Xiaorui Lyu, Haiping Peng, Dayong Wang, Haijun Yang, and Changzheng Yuan for discussions on possible invisible decay measurements at BESIII. I am also grateful to Chen-Yu Liu and the UCN$\tau$ collaboration for sharing their simulation results. Fruitful discussions with Hans P. Mumm, Kevin J. Coakley, and Paul R. Huffman on neutron lifetime experiments at NIST are appreciated as well. I thank Alexey Petrov for his comments on invisible decays via $\bar{\nu}\nu\bar{\nu}\nu$.
This work is supported in part by the National Science Foundation under grant No. PHY-1713857 and the Joint Institute for Nuclear Astrophysics (JINA-CEE, www.jinaweb.org), NSF-PFC under grant No. PHY-1430152. Funding from the faculty research support program at the University of Notre Dame is also acknowledged.
\end{acknowledgments}

\bibliography{invisible}

\end{document}